\documentclass[apj,numberedappendix]{emulateapj}

\newcommand{\be}{\begin{eqnarray}}
\newcommand{\ee}{\end{eqnarray}}
\newcommand{\lp}{\left(}
\newcommand{\rp}{\right)}

\newcommand{\E}[1]{\times10^{#1}}
\newcommand{\smpy}{ \ M_\odot \ {\rm yr}^{-1}}
\newcommand{\msol}{ \ M_\odot }
\newcommand{\commentOut}[1]{}
\newcommand{\bi}{\begin{itemize}}
\newcommand{\ei}{\end{itemize}}
\newcommand{\cgsd}{ {\rm \ g \ cm^{-3}}}
\newcommand{\cgsv}{ {\rm \ cm \ s^{-1}}}

\shorttitle{SNe .Ia: MODELS AND OBSERVABLES}
\shortauthors{SHEN, KASEN, WEINBERG, BILDSTEN, \& SCANNAPIECO}

\begin{document}


\title{Thermonuclear .Ia Supernovae from Helium Shell Detonations:\\Explosion Models and Observables}

\author{Ken J. Shen\altaffilmark{1}, Dan Kasen\altaffilmark{2,3}, Nevin N. Weinberg\altaffilmark{4}, Lars Bildsten\altaffilmark{1,5}, and Evan Scannapieco\altaffilmark{6}}
\altaffiltext{1}{Department of Physics, Broida Hall, University of California, Santa Barbara, CA 93106.}
\altaffiltext{2}{Department of Astronomy and Astrophysics, University of California, Santa Cruz, CA 95064.}
\altaffiltext{3}{Hubble Fellow.}
\altaffiltext{4}{Astronomy Department and Theoretical Astrophysics Center, 601 Campbell Hall, University of California, Berkeley, CA 94720.}
\altaffiltext{5}{Kavli Institute for Theoretical Physics, Kohn Hall, University of California, Santa Barbara, CA 93106.}
\altaffiltext{6}{School of Earth and Space Exploration, Arizona State University, P.O. Box 871404, Tempe, AZ, 85287-1404.}


\begin{abstract}

During the early evolution of an AM CVn system, helium is accreted onto the surface of a white dwarf under conditions suitable for unstable thermonuclear ignition.  The turbulent motions induced by the convective burning phase in the He envelope become strong enough to influence the propagation of burning fronts and may result in the onset of a detonation.  Such an outcome would yield radioactive isotopes and a faint rapidly rising thermonuclear ``.Ia'' supernova.  In this paper, we present hydrodynamic explosion models and observable outcomes of these He shell detonations for a range of initial core and envelope masses.  The peak UVOIR bolometric luminosities range by a factor of 10 (from $5\E{41}-5\E{42}$ erg s$^{-1}$), and the R-band peak varies from ${\rm M}_{\rm R,peak}=-15$ to $-18$.  The rise times in all bands are very rapid ($<10$ d), but the decline rate is slower in the red than the blue due to a secondary near-IR brightening.  The nucleosynthesis primarily yields heavy $\alpha$-chain elements ($^{40}$Ca through $^{56}$Ni) and unburnt He.  Thus, the spectra around peak light lack signs of intermediate mass elements and are dominated by Ca \textsc{ii} and Ti \textsc{ii} features, with the caveat that our radiative transfer code does not include the non-thermal effects necessary to produce He features.

\end{abstract}

\keywords{binaries: close--- 
novae, cataclysmic variables---
nuclear reactions, nucleosynthesis, abundances---
supernovae: general---
white dwarfs}


\section{Introduction}

AM Canum Venaticorum systems (AM CVns) are He-transferring binaries with orbital periods of $5-60$ min consisting of a degenerate or semi-degenerate He donor and a C/O or O/Ne white dwarf (WD) accretor \citep{warn95,nele05a}.  There are $\simeq 20$ known sources, many discovered recently with the Sloan Digital Sky Survey \citep{york00,ande05,roel05,roel07b,roel09,rau10}.  There are several proposed progenitor scenarios for these systems \citep{pacz67,it91,phr03}; however, recent work on the composition of the donors suggests that many, if not most, of the known AM CVns had degenerate He WD donors when mass transfer began \citep{roel09,nele09}.

In the He WD donor scenario, He mass transfer begins at rapid rates $\dot{M}=1-3\E{-6} \smpy$ and decreases with time as the binary evolves under the influence of gravitational wave radiation \citep{ty96,nele01b,db03,dbn05}.  During this initial phase of rapid accretion, the nuclear burning in the accreted He envelope is thermally stable, and the binary may appear as a supersoft X-ray source with an orbital period of a few minutes \citep{vant97,sb07}.  As $\dot{M}$ drops, the burning becomes thermally unstable, yielding $\sim 10$ He flashes in the $\sim 10^6$ yr before the amount of accreted mass required to achieve He ignition becomes larger than the available fuel remaining in the donor \citep{bild07}.  Nearly all observed AM CVns have evolved past this stage and are now slowly transferring He without any nuclear-powered phenomena.\footnote{The short orbital period systems RX J0806, V407 Vul, and ES Cet may still be in this He-flashing stage, but their parameters are too uncertain to say something definitive \citep{crop98,isra02,ww02,stro05,stro08,dt06}.}

The largest ignition masses achieved during the He-flashing stage of AM CVn evolution in the He WD donor scenario are $\sim 0.1 \msol$, large enough to possibly yield hydrodynamical burning and He detonations, which would be potentially observable as faint thermonuclear ``.Ia'' supernovae (SNe .Ia; \citealt{bild07,sb09b}).  In this paper, we further motivate our previous assumption of a He detonation and calculate observables for these predicted events.  In \S \ref{sec:turbcomb}, we calculate the conditions at which the convective burning phase of the He flash becomes inefficient at transporting entropy throughout the envelope.  We find that the subsequent onset of hydrodynamical burning occurs in the turbulence-dominated ``distributed'' burning regime, which has been suggested by \cite{nw97}, \cite{kow97}, and others to lead to a deflagration-to-detonation transition in Type Ia supernovae.  We thus proceed under the assumption that the He-burning yields a detonation.  In \S \ref{sec:hydro}, we describe the hydrodynamic and nucleosynthetic evolution of these He detonations under the assumption of 1D spherical symmetry.  In \S \ref{sec:rad}, we use the Monte Carlo radiative transfer code SEDONA \citep{ktn06} to model the light curves and spectra from these events.  We compare to known peculiar supernovae and conclude in \S \ref{sec:conc}.


\section{Convective to dynamic phase}
\label{sec:turbcomb}

During the accretion phase of the He flash cycle, radiative diffusion efficiently transports the entropy that leaks out of fluid elements within the accreted layer.  However, when nuclear reactions at the base of the envelope become significant, radiative diffusion is unable to match the increase in entropy generation, and convection begins.  The hydrostatic evolution of the growing convective zone \citep{fs82}, as recently outlined in \cite{sb09b}, is long enough that spherical symmetry is a good assumption.

The convective burning phase continues until convection becomes inefficient at transporting the energy generated by triple-$\alpha$ burning at the base of the envelope throughout the whole convective region.  For an analogous analysis in Type Ia supernovae (SNe Ia), see, e.g., \cite{hn00}, \cite{wwk04}, \cite{kwg06}, and \cite{pc08}.  This transition occurs when the local heating timescale at the base of the convective zone, $t_{\rm heat} \equiv c_P T_b / \epsilon_b$, becomes shorter than the convective eddy turnover timescale, $t_{\rm eddy} \equiv H/v_c$, after which the evolution of the burning layer effectively decouples from the rest of the convective region.  Here $c_P=3kT_b/8m_p$ is the specific heat at constant pressure in electron degenerate He, $b$-subscripts refer to the envelope base, $\epsilon_b$ is the energy generation rate, $H=P_b/\rho_b g$ is the pressure scale height, and $g$ is the local gravitational acceleration.  In mixing-length theory, the convective velocity, $v_c$, can be expressed via the convective flux as $F_c \sim \epsilon_b \rho_b H \sim \rho_b v_c^3$, so that $v_c \sim (\epsilon_b H)^{1/3}$.  When evaluated at $t_{\rm heat} = t_{\rm eddy}$, the convective velocity in electron degenerate He is $v_c' \sim (c_P T_b)^{1/2} \sim 2\E{8} \cgsv \ T_9^{1/2}$, where $T_9$ is the base temperature in units of $10^9$ K at the onset of dynamic burning.  This velocity is roughly equivalent to the thermal speed of the ions, but both are still subsonic because the sound speed is determined by the degenerate electrons and is a factor of $(E_F/kT)^{1/2}$ faster, where $E_F$ is the Fermi energy; see column 4 of Table \ref{tab:nucleo} for the initial degree of degeneracy for our models.

This large scale convective motion yields a cascade of turbulent velocities to shorter lengthscales, $l$; if we assume Kolmogorov scaling, these velocities are given by $v_{\rm turb}(l) \sim v_c (l/H)^{1/3}$.  After the transition to inefficient convection, the bulk convective motion and its associated turbulent cascade is effectively frozen out with respect to the evolution of the burning layers.  Thus, there is a spectrum of turbulent velocities present at the initiation of any burning wave in the convective shell, independent of the buoyancy-driven turbulence associated with rising flames.\footnote{That said, the convectively-driven turbulent cascade does not differ strongly from the cascade driven by a buoyant flame.  The buoyancy speed of a laminar He flame is shown in Figure 2 of \cite{tn00} to be $3\E{7} - 10^8 \cgsv$, which is roughly equal to $v_c' =8\E{7} \cgsv$, calculated at $T_b = 2\E{8}$ K, and since the lengthscale for both processes is $\sim H$, the turbulent cascades are nearly the same.}

The presence of these turbulent motions will dramatically modify the propagation of a laminar deflagration wave if the turbulent velocity at the lengthscale of the laminar flame width exceeds the burning speed.  Presuming just the turbulence generated by buoyancy (but see footnote 1), studies of laminar He deflagrations \citep{timm00,tn00} have found that turbulence indeed dominates the flame properties for initial  $\rho \lesssim 10^7 \cgsd$.  This turbulence-dominated condition is often referred to as the ``distributed'' burning regime \citep{pope87,nw97,pete00,aspd08}.  AM CVn evolution yields convective burning envelopes with densities of $ \le 3\E{6} \cgsd$ \citep{bild07,sb09b} and the prolonged convective phase ensures a turbulent medium, so that deflagrations in these He shells will be in the distributed burning regime.

In the SN Ia literature, it has been suggested that a deflagration in the distributed regime can subsequently transition to a detonation via the Zel'dovich gradient mechanism \citep{zeld70,bk86,bk87,kow97,nk97,nw97,seit09}.  One criterion for the transition to a detonation is that the large scale turbulent velocities should not be too subsonic ($\gtrsim 0.2 c_s$; \citealt{woos07,woos09}).  For a convectively-driven turbulent spectrum fixed approximately at the point that convection becomes inefficient, the convective velocity is $v_c' \sim 0.3 c_s$ (a factor of $\sqrt{kT/E_F}$ slower than the sound speed as discussed previously), which meets this criterion.  The exact outcome of distributed burning in SNe Ia is still unclear \citep{niem99}, and the conclusions are even muddier in the case of less well-studied He deflagrations, but it is plausible that a detonation will result from convectively-inefficient burning in AM CVn systems; for the rest of this study, we will examine the outcome of such shock-driven burning, with the caveat that a He deflagration may yield a qualitatively different outcome.

We will further assume a spherically symmetric geometry for the system even though multi-dimensional effects will be important for the explosion.  In particular, the detonation will likely be initiated at a single point and spread tangentially around the WD with quasi-cylindrical symmetry, as explored by \cite{fhr07,fink10}.  We leave this complication for future work.


\section{Hydrodynamic evolution and nucleosynthesis}
\label{sec:hydro}

In \cite{bild07} and \cite{sb09b}, we performed preliminary calculations for the initial conditions and general appearance of these He shell detonations.  We now proceed to probe the range of observable SN .Ia outcomes.  We evolve six combinations of core and envelope masses (core + envelopes masses: $0.6+0.2$, $0.6+0.3$, $1.0+0.05$, $1.0+0.1$, $1.2+0.02$, and $1.2+0.05 \msol$) from the onset of hydrodynamical burning.  Note that envelope masses $>0.1 \msol$ are not expected to be achieved during the evolution of AM CVn systems in the He WD donor scenario; however, we include them in our models to allow for their existence in the He-burning star donor scenario (\citealt{it91}; see, e.g., Fig. 1 of \citealt{sb09b}).  The lower of the two envelope masses for each core mass is chosen to be near the minimum envelope mass for hydrodynamical burning \citep{bild07,sb09b}, and the higher envelope's mass is chosen to be roughly twice as large.

The isothermal $T=10^7$ K cores are composed of $50\%$ $^{12}$C and $50\%$ $^{16}$O by mass, and the initially hydrostatic and isentropic envelopes are composed of pure He.  The initial thermodynamic conditions at the envelope base are set near the point when burning becomes hydrodynamic, as described in \S \ref{sec:turbcomb}.  This initial configuration is evolved with a 1D spherically symmetric explicit Lagrangian hydrodynamic code (see, e.g., \citealt{benz91}), as described in \cite{wb07}.  The 13 isotope $\alpha$-chain nuclear network\footnote{http://www.cococubed.com/code\_pages/net\_aprox13.shtml and \cite{timm99}} includes $\alpha$-capture, heavy-ion, and $(\alpha,p)(p,\alpha)$ reactions.  Different initial compositions \citep{sb09b} and neglected nuclear reactions could change the nucleosynthetic outcome of these models; we leave an exploration of these effects for future work.

\begin{figure}
	\plotone{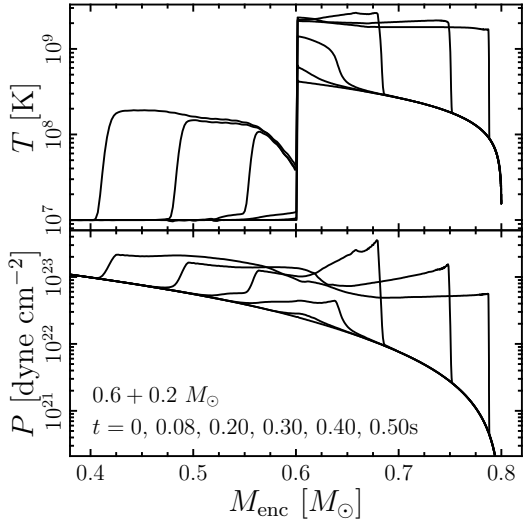}
	\caption{Snapshots of the temperature profile (\emph{top panel}) and pressure profile (\emph{bottom panel}) of the $0.6+0.2 \msol$ model during its hydrodynamic evolution.  Solid lines show the temperature and pressure vs.$ \ $the enclosed mass at 0, 0.08, 0.2, 0.3, 0.4, and 0.5 s after the beginning of the hydrodynamic simulation when the temperature at the base of the envelope is perturbed upward by $\sim 10\%$.}
	\label{fig:shockevol}
\end{figure}

At the beginning of the hydrodynamic simulation, the temperature at the base of the envelope is perturbed upwards by $\sim 10\%$, which results in a thermonuclear runaway and a pressure perturbation that steepens into a shock and an outwardly propagating detonation.  Figure \ref{fig:shockevol} shows the evolution of the temperature and pressure profiles for the $0.6+0.2 \msol$ model during this phase.  The solid lines mark the temperature and pressure vs.$ \ $the enclosed mass (i.e., the surface is to the right) at 0, 0.08, 0.2, 0.3, 0.4, and 0.5 s after the beginning of the hydrodynamic simulation.

\begin{figure}
	\plotone{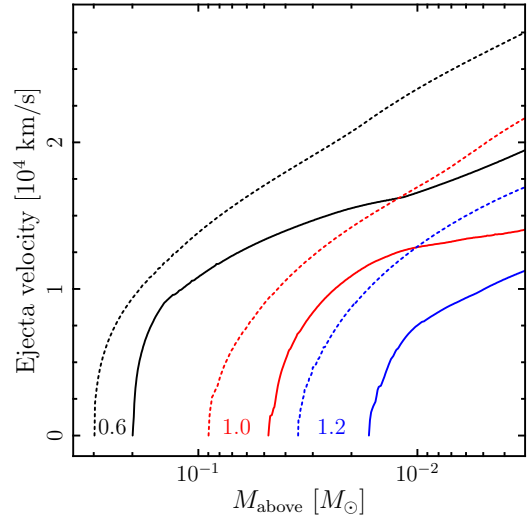}
	\caption{Free-streaming ejecta velocity vs.$ \ $mass above for all models: $0.6+0.2$, $0.6+0.3$, $1.0+0.05$, $1.0+0.1$, $1.2+0.02$, and $1.2+0.05 \msol$.  Core masses are as labeled; black, red, and blue lines are for core masses of $0.6$, $1.0$, and $1.2 \msol$, respectively.  The solid lines are the smaller envelope masses for each model, and the dashed lines are the larger envelope masses.  The velocities shown are calculated by taking the velocities 100 s after shock breakout and subtracting the effect of leaving the gravitational potential wells.}
	\label{fig:vejs}
\end{figure}

After the shock reaches the surface of the envelope, nuclear reactions are turned off for computational simplicity (as the post-shock breakout nucleosynthesis was found to be insignificant in a test case run with FLASH\footnote{http://flash.uchicago.edu}; \citealt{fryx00}).  The core and envelope are further evolved until the ejected material reaches a free-streaming, homologously expanding state.  Figure \ref{fig:vejs} shows the free-streaming ejecta velocity vs.$ \ $mass above (i.e., $M_{\rm above} = M_{\rm total}-M_{\rm enc}$) for all models: $0.6+0.2$, $0.6+0.3$, $1.0+0.05$, $1.0+0.1$, $1.2+0.02$, and $1.2+0.05 \msol$.  The velocities shown are the velocities 100 s after shock breakout with a small correction to account for leaving the gravitational potential well: i.e., $v_{\rm ej}^2 = v^2-  2 G M_{\rm enc} / r $, where the velocity, $v$, and radius, $r$, of the fluid elements are evaluated at 100 s after explosion.  We neglect any impact of the binary motion.  For larger core masses and thus larger initial gravitational potential wells, less of the envelope is able to escape, so that nearly 100\% of the $0.2 \msol$ envelope on the $0.6 \msol$ core is ejected, whereas only $\simeq 75\%$ of the $0.05 \msol$ envelope on the $1.2 \msol$ core is ejected.

\begin{table}
	\begin{center}
	\caption{Ejecta parameters}
	\begin{tabular}{|c|c||c|c|c|c|}
	\hline
	$M_{\rm WD}$ & $M_{\rm env}$ & $M_{\rm ej}$ & $E_{\rm kin,\infty}$ & $ \langle v_{\rm ej}^2 \rangle^{1/2}$ & $t_{\rm dur}$ \\
	$[M_\odot]$ & $[M_\odot]$ & $[M_\odot]$ & [$10^{50}$ erg] & [$10^4$ km s$^{-1}$] & [d] \\
	\hline
	\hline
	0.6 & 0.2 & 0.20 & 2.7 & 1.2 & 6.6 \\
	\hline
	0.6 & 0.3 & 0.30 & 5.3 & 1.3 & 7.6 \\
	\hline
	1.0 & 0.05 & 0.048 & 0.53 & 1.1 & 3.4 \\
	\hline
	1.0 & 0.1 & 0.090 & 1.2 & 1.2 & 4.4 \\
	\hline
	1.2 & 0.02 & 0.017 & 0.13 & 0.90 & 2.2 \\
	\hline
	1.2 & 0.05 & 0.036 & 0.43 & 1.1 & 2.9 \\
	\hline
	\end{tabular}
	\label{tab:deriv}
	\end{center}
\end{table}

The ejecta parameters for each model are summarized in Table \ref{tab:deriv}, which shows the ejecta mass, $M_{\rm ej}$ (column 3), total integrated kinetic energy in the ejecta, $ E_{\rm kin,\infty} \equiv \int (v_{\rm ej}^2/2) dm$ (column 4), average ejecta velocity, $\langle v_{\rm ej}^2 \rangle^{1/2} \equiv \sqrt{ 2E_{\rm kin,\infty}/M_{\rm ej} }$ (column 5), and light curve duration, $t_{\rm dur} \equiv \sqrt{ \kappa M_{\rm ej} / 7 c \langle v_{\rm ej} \rangle}$, with assumed opacity $\kappa = 0.2 $ cm$^2$ g$^{-1}$ for a He-rich medium (column 6; see \S \ref{sec:rad} for more details).  The average ejecta velocities are all very similar, between $0.9 - 1.3\E{4}$ km s$^{-1}$.  This results from the similar mass fractions of burned material in each model as well as the fact that burning from $^4$He to $^{32}$S releases only 10\% less energy per mass as burning all the way to $^{56}$Ni ($1.4\E{18}$ vs.$ \ 1.5\E{18}$ erg g$^{-1}$); see \S \ref{sec:hydro} and Table \ref{tab:nucleo} for more details.

As evident in Figure \ref{fig:shockevol}, the strong pressure increase due to the hydrodynamic burning in the envelope sends a shock wave into the WD core.  A sequence of events like this has been discussed extensively in the context of double degenerate Type Ia supernova progenitors, either as an edge-lit detonation in which the shock ignites C-burning at the core-envelope interface \citep{nomo82b,lg90,lg91,la95,wsf98,gbw99}, or as a shock-focused detonation in which the geometrically-focused shock yields a carbon detonation near the center of the WD \citep{livn90,lg91,ww94,la95,fhr07,fink10,sim10}.  In our models, where the temperature is only perturbed at the exact interface between the envelope and core, the inwardly propagating shock is not strong enough to immediately ignite the $^{12}$C at the core's edge.  For detonations initiated well above the base of the envelope, the inwardly propagating shock strengthens as it approaches the envelope-core interface and may ignite the edge of the core \citep{lg90,gbw99}; we do not consider this possibility.  Given the uncertainties in these detonation models, and the expectation that some fraction of the AM CVn population contains O/Ne WD accretors, which would be much more difficult to detonate via these mechanisms, we proceed under the assumption that the core is not detonated, with the caveat that this outcome remains a possibility.


\begin{table*}
	\begin{center}
	\caption{Ejecta composition}
	\begin{tabular}{|c|c|c|c||c|c|c|c|c|c|c|c|c|}
	\hline
	$M_{\rm WD}$ & $M_{\rm env}$ & $\rho_b$ & $kT_b/E_F$ & $M_{\rm ej}$ & $X_4$ & $X_{36}$ & $X_{40}$ & $X_{44}$ & $X_{48}$ & $X_{52}$ & $X_{56}$ & $\rho_{\rm eq}$ \\
	$[M_\odot]$ & $[M_\odot]$ & [$10^5$ g cm$^{-3}$] & & $[M_\odot]$ & & &  &  &  & & & [$10^5$ g cm$^{-3}$]  \\
	\hline
	\hline
	0.6 & 0.2 & 6.8 & 0.28 & 0.20 & 0.45 & 0.0067 & 0.036 & 0.12 & 0.13 & 0.12 & 0.13 & 4.1 \\
	\hline
	0.6 & 0.3 & 24 & 0.083 & 0.30 & 0.30 & $1.2\E{-4}$ & 0.0020 & 0.0072 & 0.0097 & 0.022 & 0.65 & 3.6 \\
	\hline
	1.0 & 0.05 & 5.8 & 0.40 & 0.048 & 0.44 & 0.034 & 0.30 & 0.11 & 0.032 & 0.033 & 0.015 & $-$ \\
	\hline
	1.0 & 0.1 & 21 & 0.10 & 0.090 & 0.39 & $5.8\E{-5}$ & 0.0014 & 0.0066 & 0.011 & 0.032 & 0.56 & 5.1 \\
	\hline
	1.2 & 0.02 & 9.3 & 0.28 & 0.017 & 0.51 & 0.0037 & 0.025 & 0.063 & 0.081 & 0.17 & 0.13 & 5.8 \\
	\hline
	1.2 & 0.05 & 34 & 0.068 & 0.036 & 0.38 & $3.6\E{-5}$ & $8.0\E{-4}$ & 0.0033 & 0.0052 & 0.015 & 0.59 & 6.6 \\
	\hline
	\end{tabular}
	\label{tab:nucleo}
	\end{center}
\end{table*}

\begin{figure}
	\plotone{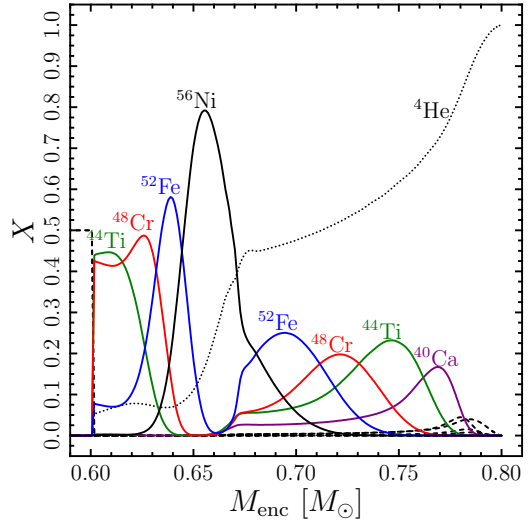}
	\caption{Mass fractions vs.$ \ $enclosed mass at the time of shock breakout for the $0.6+0.2 \msol$ model.  Solid lines are for the isotopes as labeled, and dashed lines are for the remaining isotopes, which are insignificant in the ejecta.}
	\label{fig:comp}
\end{figure}

Table \ref{tab:nucleo} shows the ejecta mass (column 5) for each model, denoted by its core and envelope masses (columns 1 \& 2), initial base density (column 3), and initial degree of degeneracy (column 4), along with the mass fractions of $^4$He, $^{36}$Ar, $^{40}$Ca, $^{44}$Ti, $^{48}$Cr, $^{52}$Fe, and $^{56}$Ni in the ejecta (columns 6-12).  Column 13 shows $\rho_{\rm eq}$, the density where the He and Ni mass fractions are equal, except for the $1.0+0.05 \msol$ model for which the two mass fractions were not equal anywhere in the envelope.  For the larger envelope masses for each core mass, the majority of burning products in the He shell are in the form of $^{56}$Ni, with decreasing contributions from $^{52}$Fe, $^{48}$Cr, and $^{44}$Ti.  For the smaller envelope masses, the contribution of isotopes below $^{56}$Ni are enhanced as compared to the larger envelope masses.  In particular, the $1.0+0.05 \msol$ model, with a relatively low initial base density $ \simeq 6\E{5} \cgsd$, only produced a small amount of $^{56}$Ni.  Its nucleosynthesis is dominated by the production of $^{40}$Ca.

Profiles of mass fraction vs.$ \ $ejecta mass coordinate are shown in Figure \ref{fig:comp} for a representative model ($0.6+0.2 \msol$).  Solid lines are for the isotopes as labeled, and dashed lines are for the remaining isotopes, which are insignificant in the ejecta.  Near the base of the initial He shell, where the fluid elements are confined at high temperatures and densities for the longest time, the nucleosynthesis is dominated by Fe-group elements at the end of the $\alpha$-chain.  Closer to the surface, where there is little time to burn the fuel prior to radial expansion, most of the He remains unburnt.  For the densities and temperature reached in these detonations, the slowest step of He-burning is the first: triple-$\alpha$ burning.  If there is time to form a $^{12}$C nucleus, it will capture additional He nuclei relatively rapidly and proceed all the way to Fe-group elements near the end of the $\alpha$-chain.  Thus, for all the models, very few intermediate mass elements are produced; the cumulative mass fraction of isotopes between $^{20}$Ne and $^{32}$S for each model is $<1\%$.  \cite{pere09}'s one-zone analysis produces a qualitatively similar nucleosynthetic yield for the relevant temperatures and their most He-rich composition.  This has strong implications for the spectral signatures of these models, which will be discussed in \S \ref{sec:spec}.


\section{Radiative transfer and observables}
\label{sec:rad}

After reaching the free-streaming phase, the hydrodynamic output described in \S \ref{sec:hydro} is processed by SEDONA, a Monte Carlo radiative transfer code \citep{ktn06}. The necessary modifications to follow the radioactive decays of $^{48}$Cr and $^{52}$Fe have been included as outlined in \cite{bild07}.

We can gain a qualitative estimate of the light curve properties by employing the simplifying assumptions derived in \cite{arne82} and \cite{pe00a}: namely, that the duration of the light curve, $t_{\rm dur}$, has a scale set primarily by the time at which the radiative diffusion time equals the age of the explosion (column 6 of Table \ref{tab:deriv}), and that the peak bolometric luminosity, $L_{\rm peak}$, is equal to the instantaneous radioactive decay at $t=t_{\rm dur}$.\footnote{The low mass envelopes remain radiation dominated throughout their evolution.}  These basic assumptions yield
\be
	t_{\rm dur} \simeq  5 {\rm \ d } \left[ \lp \frac{\kappa}{0.2 {\rm \ cm^2 \ g^{-1} } } \rp \lp \frac{M_{\rm ej}}{0.1 \msol } \rp \lp \frac{10^4 {\rm \ km \ s^{-1} } }{v_{\rm ej}} \rp \right]^{1/2}
\ee
and
\be
	L_{\rm peak} \simeq \frac{Q_i N_i}{\tau_i} \exp \lp - \frac{t_{\rm dur}}{\tau_i} \rp ,
	\label{eq:lpeak}
\ee
where the $i$-subscript refers to the dominant radioactive product in the ejecta of which there are $N_i$ nuclei, $Q_i$ is the energy release per radioactive decay, and $\tau_i$ is the $e$-folding lifetime for the decay.  This approximation of the peak luminosity yields $L_{\rm peak} \sim 2\E{42}$ erg s$^{-1}$ or a peak bolometric magnitude ${\rm M}_{\rm bol} \sim -17$ for $0.05 \msol$ of $^{56}$Ni and $t_{\rm dur} = 5$ d.  Note that this estimate assumes that the ejecta is optically thick to all radiation, including $\gamma$-rays, at peak.  This estimate is further complicated by the inclusion of multiple radioactive burning products, each with more than one step in their decays to their stable end states.  Additionally, these radioactive isotopes have different diffusion times due to their vertical stratification in the envelope, as shown in Figure \ref{fig:comp}.  However, equation (\ref{eq:lpeak}) is useful in describing general trends in $L_{\rm peak}$.  In particular, it demonstrates two offsetting features of SNe .Ia: while the low radioactive mass (i.e., small $N_i$) lowers $L_{\rm peak}$ as compared to SNe Ia, the low ejecta mass implies a short $t_{\rm dur}$, thus increasing $L_{\rm peak}$.  The net effect is to decrease $L_{\rm peak}$, but not as much as one might na\"{i}vely assume given the amount of radioactive material.


\subsection{Light curves}

\begin{figure}
	\plotone{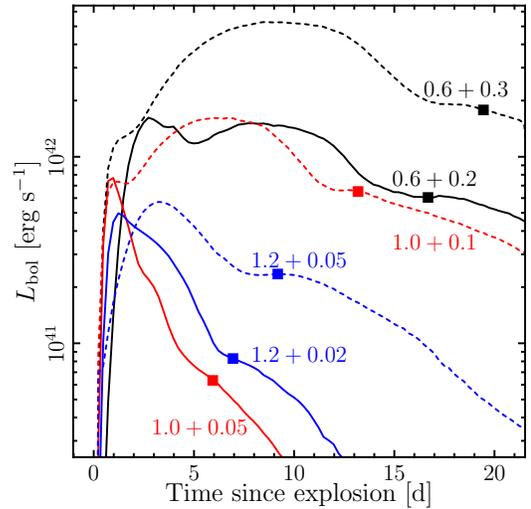}
	\caption{UVOIR bolometric luminosity vs.$ \ $time since explosion for all models as labeled.  Solid lines are for the lower envelope mass for each core, and dashed lines are for the higher envelope mass.  Black, red, and blue lines are for core masses of $0.6$, $1.0$, and $1.2 \msol$, respectively.  The multiple peaks are due to the stratification of radioactive isotopes in the ejecta.  Squares denote the time after which over half of the gamma-rays escape the ejecta without being absorbed.}
	\label{fig:lcs}
\end{figure}

Figure \ref{fig:lcs} shows UVOIR bolometric light curves for the six models as labeled.  Solid lines are for the lower  envelope mass for each core, and dashed lines are for the higher envelope mass.  Black, red, and blue lines are for core masses of $0.6$, $1.0$, and $1.2 \msol$, respectively.  Squares denote the time after which over half of the gamma-rays escape the ejecta without being absorbed.  As expected from the above discussion and \cite{bild07}, the light curves reach peak luminosity fairly rapidly, within $2-10$ d after the detonation, with peak luminosities in accord with the energy input from instantaneous radioactive decays.  The peak luminosities range over an order of magnitude from $5\E{41}-5\E{42}$ erg s$^{-1}$, corresponding to peak bolometric magnitudes of $-15.5$ to $-18$.

Radioactive energy deposited in mass shells closer to the surface have shorter diffusion timescales.  Thus, the stratification of radioactive isotopes in the ejecta, as demonstrated in Figure \ref{fig:comp} for the $0.6+0.2$ model, yields multiple peaks in the bolometric light curves of models with envelope masses $ \geq 0.1 \msol$ for which the light curves are broad enough to distinguish the multiple peaks.  In these models, the first peak or shoulder is due to the rapid decay of $^{48}$Cr and $^{52}$Fe near the surface, followed by a broader peak due to the decay of these isotopes and $^{56}$Ni deeper within the ejecta.

\begin{figure}
	\plotone{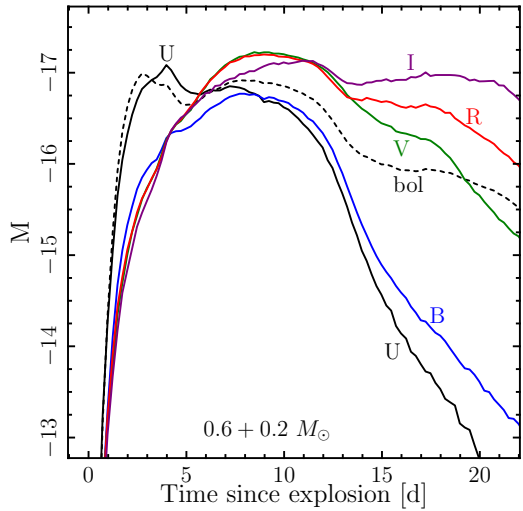}
	\caption{Multiband light curves for the $0.6+0.2 \msol$ model in Vega magnitudes.  The dashed line shows the bolometric light curve, and the solid lines show magnitudes for various filters as labeled.  The secondary peak in redder wavebands is apparent beginning $\simeq 15$ d after explosion, which is $\simeq 7$ d after the time of B-band maximum.}
	\label{fig:0602color}
\end{figure}

The Vega magnitudes in the UBVRI filters vs.$ \ $time since explosion for the $0.6+0.2 \msol $ model are shown as solid lines as labeled in Figure \ref{fig:0602color}; the magnitude of the bolometric light curves is shown as a dashed line.  The early peak in the bolometric light curve due to the stratification of isotopes $\simeq 2$ d after the explosion yields an early peak in the U filter, since the luminosity, radius, and the Wien displacement law give a maximum in the spectral energy at $\simeq 2000$ \AA \ at this time.  The second bolometric peak, primarily powered by the decay of $^{56}$Ni, yields maxima in the other filters at roughly the same time as each other, $\simeq 8$ d after explosion.

\begin{figure}
	\plotone{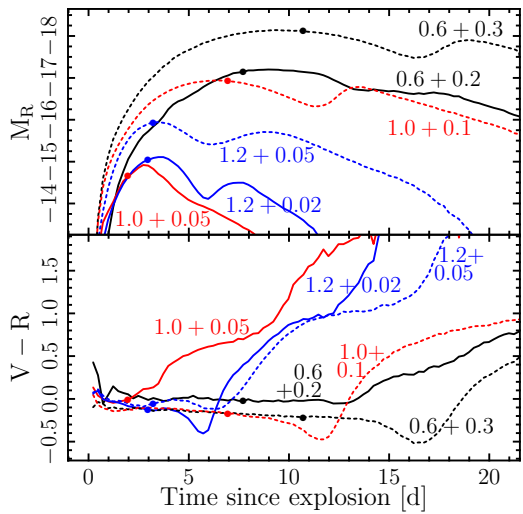}
	\caption{R-band light curves (\emph{top panel}) and ${\rm V}-{\rm R}$ colors (\emph{bottom panel}) for all models as labeled.  Solid lines show light curves and colors for the smaller envelope for each core mass, and dashed lines represent the larger envelopes.  Black, red, and blue lines are for core masses of $0.6$, $1.0$, and $1.2 \msol$, respectively.  Bullets denote the time of the B-band maximum for each model.}
	\label{fig:randcolor}
\end{figure}

Figure \ref{fig:randcolor} shows the R-band light curves (\emph{top panel}) and ${\rm V}-{\rm R}$ colors (\emph{bottom panel}) for all models.  Bullets mark the time of the B-band maximum for each model.  All of the models have ${\rm V}-{\rm R} = -0.2 $ to $0$ at the time of B-band maximum, and all then become much redder at later times, with most possessing a secondary peak in the R, I, and redder bands.  For the $0.6+0.2 \msol$ model shown in Figure \ref{fig:0602color}, for example, the B-band light curve takes 6 d to drop 1 mag from B-band peak, while it takes 12 d for the R-band light curve to drop 1 mag from its peak.  This strong reddening is due to the same physics that produces secondary near-IR peaks in typical SNe Ia: namely, the emissivity of Fe-group isotopes peaks sharply in near-IR filters when they transition from the doubly- to singly-ionized state at $\simeq 7000$ K \citep{pe00b,kase06}.


\subsection{Spectra}
\label{sec:spec}

\begin{figure}
	\plotone{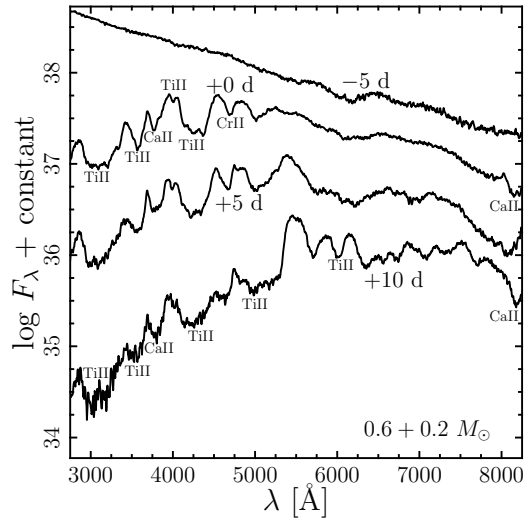}
	\caption{Optical spectra for the $0.6+0.2 \msol$ model at $-5$, $+0$, $+5$, and $+10$ d with respect to B-band maximum, which occurs 8 d after explosion.  The spectra are dominated by Ti \textsc{ii} features, along with the Ca \textsc{ii} H\&K and IR lines and a short-lived Cr \textsc{ii} feature near 4600 \AA.  The lack of Si \textsc{ii} features is also a notable signature.}
	\label{fig:specevol}
\end{figure}

The optical spectra of the $0.6+0.2 \msol$ model are shown in Figure \ref{fig:specevol} at $-5$, $+0$, $+5$, and $+10$ d with respect to the B-band maximum, which occurs 8 d after the explosion.  The spectra are offset by arbitrary constants.  Line and feature identifications in the spectra are made by comparing the complete spectra to radiative transfer simulations in which the effect of a single isotope is removed.

The pre-maximum spectrum is a fairly featureless blue continuum.  By the time of B-band maximum, Ti \textsc{ii} features have become prominent.  The Ca \textsc{ii} H\&K lines and IR triplet are also present with blueshifts of $1.4\E{9} \cgsv$, as is a Cr \textsc{ii} feature near 4700 \AA \ that is likely unique to supernovae powered by He-burning products.  At 10 d after the time of B-band maximum (18 d after explosion), the continuum has become much redder due to the reprocessing of UV and blue light by Fe-group elements, as previously discussed.  The Ca \textsc{ii} lines are still visible, but the spectrum is determined primarily by Ti \textsc{ii} features.  The Cr feature at 4700 \AA \ has disappeared by this point due to $^{48}$Cr's short 22 hr half-life.

The spectra are also notable in what they do not show.  First, they lack lines from intermediate mass elements such as Si, S, and Mg.  This is not surprising given the negligible amount of these isotopes produced in these models.  Second, the spectra show no signs of He, which may seem counterintuitive given the large amount of unburnt He in the ejecta.  However, the relevant optical He lines have lower levels with high excitation energies ($>10$ eV) and will not be highly populated by thermal processes given the moderate ejecta temperatures $\sim 10^4$ K.  As the present calculations assume local thermodynamic equilibrium (LTE) to calculate the level populations, none of the optical He lines has significant opacity.  However, the gamma-rays from radioactive decay will Compton scatter and produce fast electrons that may non-thermally excite these levels and thereby increase the line opacity.  Inclusion of this non-LTE effect is necessary to explain the He lines observed in Type Ib supernovae \citep{lucy91}.

The strengths of the He lines in supernovae are therefore not only sensitive to the He mass, but also to the amount of radioactive material present and on how this material is mixed into the He layer.  In the SN .Ia case, radioactive material is produced adjacent to the He and the densities are relatively low, so gamma-rays do effectively diffuse into the He layer.  It therefore seems likely that non-thermal excitation will produce He lines in some models, but we need quantitative non-LTE calculations in the future to confirm this.

\begin{figure}
	\plotone{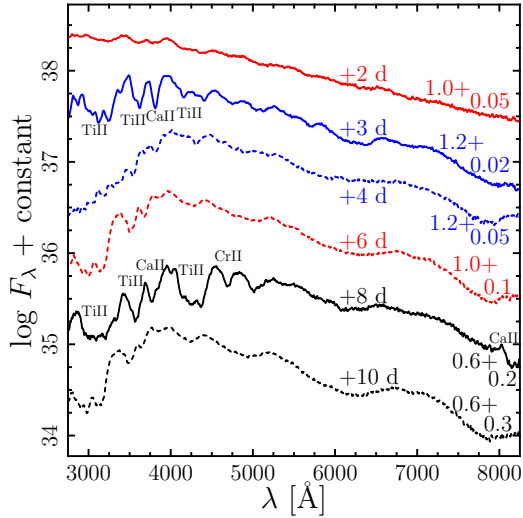}
	\caption{Optical spectra at B-band maximum for all models as labeled, with arbitrary offsets.  Black, red, and blue lines are for 0.6, 1.0, and $1.2 \msol$ cores, respectively.  Solid lines are for the lower envelope mass for each model, and dashed lines are for the higher envelope mass.  Times shown are the ages since explosion.}
	\label{fig:specmax}
\end{figure}

Figure \ref{fig:specmax} shows optical spectra for all the models as labeled at the time of their respective B-band maxima.  Black, red, and blue lines are for 0.6, 1.0, and $1.2 \msol$ cores, respectively; solid lines are for the lower envelope mass for each model, and dashed lines are for the higher envelope mass.  The times shown are the ages since explosion for each model at their B-band maxima.  Most models show Ca \textsc{ii} and Ti \textsc{ii} absorption lines with varying strengths depending on the nucleosynthetic output and the photospheric temperature, which is itself a function of the luminosity and radius at B-band maximum.  E.g., the lower envelope mass models for each core mass ($0.6+0.2$, $1.0+0.05$, and $1.2+0.02 \msol$) are the most $^{40}$Ca- and $^{44}$Ti-rich, and two of these three models reflect this; however, the B-band maximum for the $1.0+0.05 \msol$ model occurs so early that the photosphere is hot enough to produce a relatively featureless blue continuum.


\section{Conclusions}
\label{sec:conc}

Previous work by \cite{bild07} and \cite{sb09b} hypothesized that a new class of SNe .Ia powered by He shell detonations could be produced by AM CVn systems, and in this work, we have calculated their observable properties in depth for the first time.  In \S \ref{sec:turbcomb}, we examined the outcome of turbulent combustion with guidance from work done on Type Ia supernovae and found that the assumption of a He detonation is well-motivated by the initial conditions.  The hydrodynamic evolution of these detonations was calculated in 1D and led to the ejection of most of the envelopes with velocities of $\sim 10^4$ km s$^{-1}$ (Figs. \ref{fig:shockevol} and \ref{fig:vejs} and Table \ref{tab:deriv}).  The nucleosynthetic output of each envelope was dominated by unburnt He and $\alpha$-chain elements from $^{40}$Ca to $^{56}$Ni (Fig. \ref{fig:comp} and Table \ref{tab:nucleo}).  We employed a Monte Carlo radiative transfer code to produce light curves (Figs. \ref{fig:lcs}, \ref{fig:0602color}, and \ref{fig:randcolor}), which are notable for their fast rises of $2-10$ d, their multiple peaks due to the stratification of radioactive isotopes, and their strong late-time reddening due to Fe-group ionization effects.  The spectra are dominated by Ca \textsc{ii} and Ti \textsc{ii} features (Figs. \ref{fig:specevol} and \ref{fig:specmax}).

Recently, several peculiar Type I supernovae have been observed that do not fit cleanly into existing supernova categories.  SN 2002bj was discovered in archival Lick Observatory Supernova Search data by \cite{pozn10}.  Its most striking feature is its extremely rapid rise and fall, with a $>3.5$ magnitude R-band rise in 7 d to its peak at $\simeq -18.5$ in all bands, and a subsequent decline of 3 magnitudes in 15 d.  A spectrum taken 7 d after discovery shows signatures of He and possibly even V, which could be the decay product of $^{48}$Cr.  However, while the general speed of the light curve and the evidence of He-burning are in line with the SN .Ia predictions in this paper, SN 2002bj also shows evidence of intermediate mass elements and does not have Ti features or strong late-time reddening.

SN 2008ha was a faint (peak ${\rm M}_{\rm V}=-14.2$) and fast rising and declining supernova \citep{fole09,fole10,vale09} that at first glance seemed a possible SN .Ia.  However, its slow ejecta velocity of $\sim 2000$ km s$^{-1}$ and the nucleosynthetic signatures of C-burning point to the failed deflagration of a C/O WD instead of a He shell detonation.

\begin{figure}
	\plotone{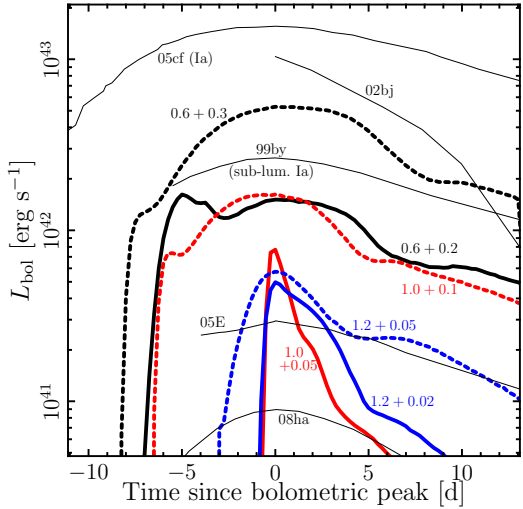}
	\caption{UVOIR bolometric luminosity vs.$ \ $time since bolometric peak.  Our .Ia models are shown as thick lines as labeled; thick solid lines show the less massive envelopes for each core mass, thick dashed lines show the more massive envelopes, and black, red, and blue lines denote $0.6$, $1.0$, and $1.2 \msol$ cores, respectively.  Thin black lines are UVOIR bolometric light curves from integrated and blackbody fits for observed supernovae as labeled.}
	\label{fig:lccomp}
\end{figure}

SN 2005E was a faint (peak ${\rm M}_{\rm B}=-14.8$) Ca-rich Type Ib supernova with a rapid rise and a fall of 1 mag per $10-15$ d \citep{pere09}.  Many of its characteristics match our general results, including a lack of intermediate mass elements, although its bolometric light curve may rise too slowly (Fig. \ref{fig:lccomp}).  The nucleosynthesis does not match exactly, as SN 2005E shows evidence for a large amount of Ca as well as some C and O, but given the nucleosynthetic outcome of our weakest explosion model ($1.0+0.05 \msol$), it is possible that a still weaker explosion would yield an ejecta composition matching that of SN 2005E.  As previously mentioned, the inclusion of a larger nuclear network and multidimensional effects will likely play a role in changing the nucleosynthetic yield; these refinements await future work.

Figure \ref{fig:lccomp} compares our model light curves to those of several observed SNe.  UVOIR bolometric light curves are plotted as in Figure \ref{fig:lcs}, but vs.$ \ $the time since bolometric peak instead of the time since explosion.  Thick lines show .Ia models as throughout this paper: thick solid lines show the less massive envelopes for each core mass, thick dashed lines show the more massive envelopes, and black, red, and blue lines denote $0.6$, $1.0$, and $1.2 \msol$ cores, respectively.  Also shown as thin black lines are quasi-bolometric light curves for the three peculiar SNe just discussed as well as SN 2005cf, a well-studied bright SN Ia \citep{wang09}, and SN 1999by, a sub-luminous SN Ia \citep{garn04,phil07}.  The light curves are constructed from integrated and blackbody fits to broadband photometry; in the case of SN 2005E, which was limited to BVRI observations, the U-band and near-IR contributions are assumed to scale as they do in SN 2005cf (S. Valenti, private communication; \citealt{past07}).\footnote{For more details of the fitting method, see \cite{vale08a}.}

It is clear that the predicted SN .Ia rise times of $2-10$ d are much faster than any other rise times, aside from the extremely fast SN 2002bj, but the peak bolometric luminosities cover a large range from those of faint peculiar SNe to sub-luminous SNe Ia, and the declines in luminosity are not strongly atypical of observed SNe.  Thus, the most robust and unique signatures of a SN .Ia will be a very fast rise to peak luminosity and the appearance of He-burning products such as Ca \textsc{ii} and Ti \textsc{ii} in the spectra.


\acknowledgments

We thank Ryan Foley, Hagai Perets, Anthony Piro, Dovi Poznanski, and Dean Townsley for helpful discussions, and Stefano Valenti for providing the bolometric light curve of SN 2005E.  This work was supported by the National Science Foundation under grants PHY 05-51164 and AST 07-07633.



\end{document}